\documentstyle[preprint,aps,epsf]{revtex}
\draft
\begin{document}
\preprint{Phys. Rev. Lett. (1997) in press.}
\title{Equation of state of asymmetric nuclear matter
and collisions of neutron-rich nuclei}
\bigskip
\author{\bf Bao-An Li$^{a}$\footnote{email: Bali@comp.tamu.edu}, 
C.M. Ko$^{a}$\footnote{email: Ko@comp.tamu.edu} and 
Zhongzhou Ren$^{b}$\footnote{email: Zyq@nju.edu.cn}}
\address{$^a$ Cyclotron Institute and Department of Physics\\ 
Texas A\&M University, College Station, TX 77843, USA\\
$^b$ Department of Physics, Nanjing University, Nanjing 210008, P.R. China}
\maketitle

\begin{quote}
The ratio of pre-equilibrium neutrons to protons from collisions of 
neutron-rich nuclei is studied as a function of their kinetic energies. 
This ratio is found to be sensitive to the density dependence of the 
nuclear symmetry energy, but is independent of the compressibility of 
symmetric nuclear matter and the in-medium nucleon-nucleon cross sections. 
The experimental measurement of this ratio thus provides a novel means 
for determining the nuclear equation of state of asymmetric nuclear matter.
\end{quote}

\newpage
Recent advance in radioactive ion beam experiments has opened up a 
new field of research in nuclear physics (for a recent review, see, e.g., 
refs. \cite{tanihata95,hansen95}). These experiments have already 
provided useful information about the structure of unstable 
nuclei far from the stability valley. However, it has not been generally
recognized that they are also useful for extracting information on the 
equation of state (EOS) of asymmetric nuclear matter, which has not been 
well determined and is important for understanding both the structure 
of unstable nuclei and the properties of neutron stars. 
In relativistic mean-field (RMF) theory, it has been recently shown that 
the nuclear symmetry energy affects significantly 
the binding energy and rms radii of neutron-rich nuclei
\cite{sumi93}. Also, the chemical composition, the evolution of lepton 
profiles, and the neutrino fluxes in neutron stars depend strongly on the
nuclear symmetry energy \cite{toki95}. More important 
but poorly known is the density dependence of the nuclear symmetry energy.
Better knowledge on this quantity is required to understand
both the matter radii of many neutron-rich isotopes, which has been 
found to increase faster than $A^{1/3}$\cite{na}, and the central 
density of neutron-rich nuclei, which is lower than that of 
stable nuclei\cite{tanihata96}. In nuclear astrophysics, the density 
dependence of nuclear symmetry energy is crucial for understanding 
the supernova explosion scenarios and the cooling mechanisms of neutron 
stars\cite{lat91,sum94}. In particular, it determines the 
equilibrium concentration of protons in a neutron star. 
If the latter is larger than a critical value of about 15\%, the direct 
URCA process can happen, and would then enhance neutrino emissions and 
the neutron star cooling rate\cite{lat91}. 

M\"uller and Serot have recently shown 
that the asymmetric EOS has quite distinct new features compared to the 
symmetric one\cite{muller}. In particular, in asymmetric nuclear matter 
the liquid-gas phase transition is second order rather than first order as in 
symmetric nuclear matter. Furthermore, the instabilities 
that produce a liquid-gas phase separation in asymmetric nuclear matter 
arise from fluctuations in the neutron/proton concentration (chemical 
instability) instead from fluctuations in the baryon density 
(mechanical instability) as in symmetric nuclear matter. The latter may be 
related to the recently observed isospin-dependence of nuclear 
multifragmentation in heavy-ion collisions at intermediate 
energies\cite{kunde96,dem96}. 

Presently available radioactive ion beam facilities and planned isospin 
laboratories provide the unique opportunity to study experimentally 
the EOS of asymmetric nuclear matter. 
Tanihata has recently proposed to extract the EOS of asymmetric 
nuclear matter by studying the properties of neutron-rich nuclei, 
such as their density distributions, radii and the thickness of the 
neutron skin\cite{tanihata96}. In this Letter, 
we shall suggest a novel approach for studying the asymmetric 
part of the nuclear EOS, i.e.,  via the ratio of the number of preequilibrium 
neutrons to that of protons $(R_{{\rm n/p}}(E_{{\rm kin}})\equiv dN_n/dN_p)$ 
from collisions of neutron-rich nuclei at intermediate energies. 
Using a transport model with explicit isospin degrees of freedom, we 
will show that the ratio $R_{{\rm n/p}}(E_{{\rm kin}})$ is sensitive 
to the density dependence of the symmetry energy, but is almost 
independent of the compressibility of symmetric nuclear matter
and the in-medium nucleon-nucleon cross sections.   

Theoretical studies (e.g., 
\cite{lat91,siemens70,baym71,laga81,wiringa88,prak88,thor94}) 
have shown that the EOS of asymmetric nuclear matter can be approximately
expressed as
\begin{equation}\label{aeos}
E(\rho,\beta)=E(\rho,\beta=0)+S(\rho)\beta^2,
\end{equation}
where $\rho=\rho_n+\rho_p$ is the baryon density;
$\beta=(\rho_n-\rho_{p})/(\rho_p+\rho_n)$ is the 
relative neutron excess; and $E(\rho,\beta=0)$ is the energy 
per particle in symmetric nuclear matter. The bulk symmetry 
energy is devoted by 
$S(\rho)\equiv E(\rho,\beta=1)-E(\rho,\beta=0)$.  Its value
$S_0\equiv S(\rho_0)$ at normal nuclear matter density is 
known to be in the range of 27-36 MeV\cite{mass}. 
In the non-relativistic Hartree-Fock 
theory (e.g., \cite{farine,pear}) and the relativistic mean field (RMF) 
theory (e.g., \cite{serot,glen82,rein88,rufa88,shar93,toki94,scha96}),       
the predicted values of $S_0$ are 27-38 MeV and 35-42 MeV, respectively. 
Also, the density dependence of the symmetry energy varies widely among
theoretical studies. 
A $\rho^{1/3}$ dependence was obtained by Siemens using the 
Bethe-Goldstone theory for asymmetric nuclear matter \cite{siemens70}, 
while the RMF theory predicts a linear dependence\cite{chin77,horo87}. 
One of the most sophisticated calculations is that of Wiringa {\it et al.} 
using the variational many-body theory\cite{wiringa88}. Different 
density dependences of $S(\rho)$ have been found depending on the 
nuclear forces used in the calculation. Typical results of these studies can 
be parameterized by\cite{prak88} 
\begin{equation}
S(\rho)=(2^{2/3}-1)\frac{3}{5}E_{F}^{0}[u^{2/3}-F(u)]+S_0F(u),
\end{equation} 
with $F(u)$ having one of the following three forms
\begin{eqnarray}\label{fu}
F_1(u)&=&\frac{2u^2}{1+u},\\\nonumber
F_2(u)&=&u,\\\nonumber
F_3(u)&=&u^{1/2},
\end{eqnarray}
where $u\equiv \rho/\rho_0$ is the reduced baryon density and $E_F^0$ is the 
Fermi energy. From eq.\ (\ref{fu}) the contribution of nuclear interactions 
to the symmetry energy density can be obtained, i.e.,
\begin{equation}
w_a(\rho,\beta)=e_a\cdot \rho F(u)\beta^2, 
\end{equation}
where $e_a\equiv [(S_0-(2^{2/3}-1)\frac{3}{5}E_F^0]$ is the 
contribution of nuclear interactions to the bulk symmetry energy at normal
nuclear matter density. The mean-field potentials for neutrons and protons 
due to the symmetry energy are then 
\begin{equation}
V^{n(p)}_{{\rm asy}}(\rho,\beta)=\partial w_a(\rho,\beta)/\partial \rho_{n(p)}.
\end{equation} 

\begin{figure}[htp]
\setlength{\epsfxsize=14truecm}
\centerline{\epsffile{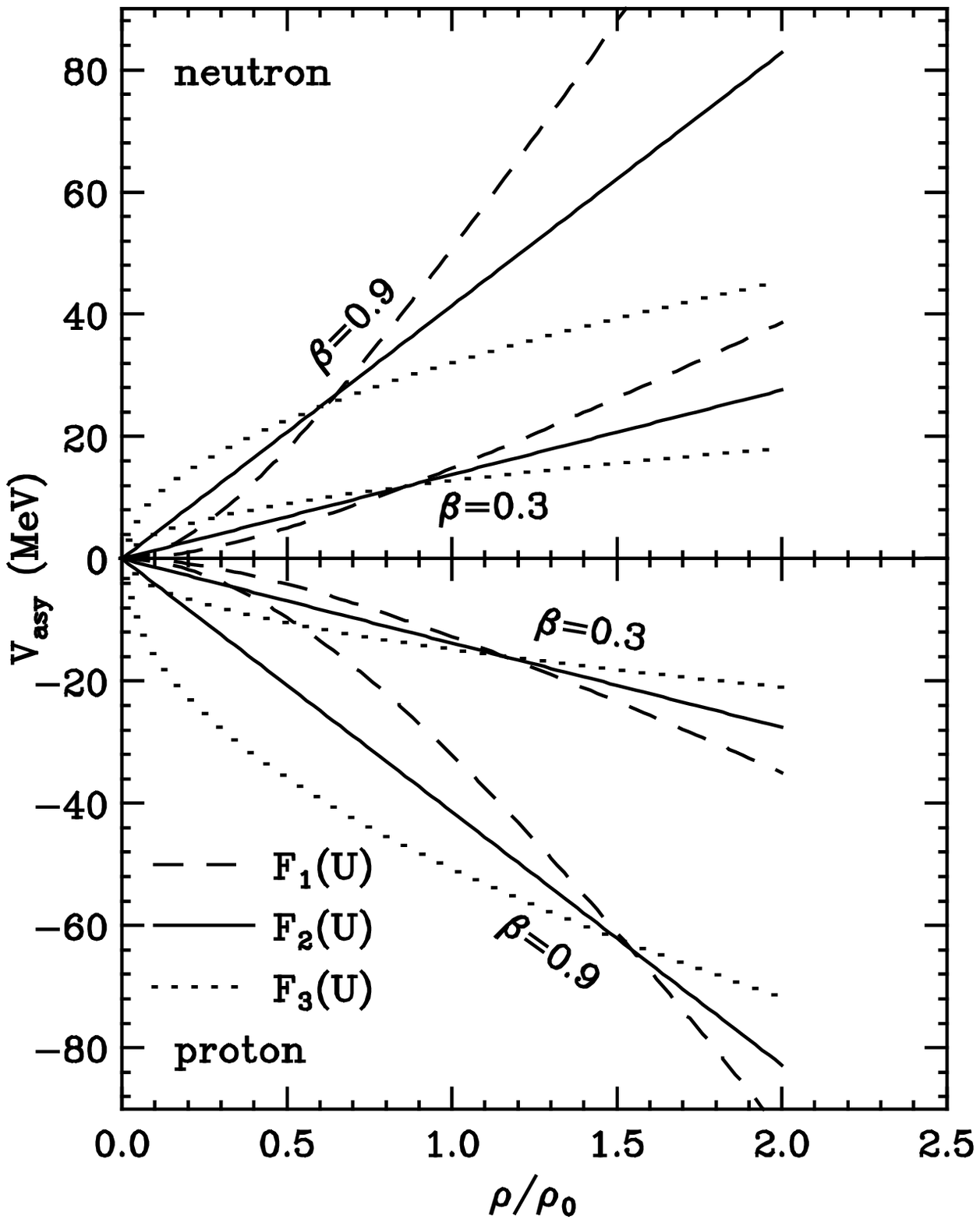}}
\caption{The symmetry potential for neutrons and protons corresponding to the
three forms of $F(u)$ (see text).}
\label{ieosfig1} 
\end{figure}

To illustrate the magnitude of the symmetry potential, we show in 
Fig.\ \ref{ieosfig1} $V_{{\rm asy}}^{n(p)}(\rho,\beta)$ using the 
three forms of $F(u)$ and $S_0=32$ MeV. 
It is seen that the repulsive (attractive) mean-field for neutrons 
(protons) depends sensitively on the form of $F(u)$, the neutron excess 
$\beta$, and the baryon density $\rho$. In collisions of neutron-rich 
nuclei at intermediate energies, both $\beta$ and $\rho$ can be appreciable
in a large space-time region where the isospin-dependent mean-fields, which are 
opposite in sign for neutrons and protons, are strong. This will affect 
differently the reaction dynamics of neutrons and protons, leading to 
possible differences in their yields and energy spectra. Since the magnitude of
the asymmetric part of the nuclear EOS is small compared to the 
symmetric part in eq.\ (\ref{aeos}), to extract $S(\rho)$ requires observables 
which are sensitive to the asymmetric part but not 
the symmetric part of the nuclear EOS. Also, these observables should not 
depend strongly on other factors that affect 
the reaction dynamics, such as the in-medium nucleon-nucleon cross sections. 
In the following, we shall demonstrate that the ratio 
$R_{n/p}(E_{{\rm kin}})$ of preequilibrium neutrons to protons from collisions
of neutron-rich nuclei meets these requirements.    

We shall use an isospin-dependent Boltzmann-Uehling-Uhlenbeck (BUU) transport 
model (e.g., \cite{greiner,bertsch,libauer1,libauer2}). The proton and 
neutron densities calculated from the nonlinear relativistic mean-field (RMF) 
theory\cite{shar93,ren} are used as inputs to initialize the BUU 
model\cite{lis95,li96}. The isospin dependence is included in the dynamics
through nucleon-nucleon collisions by using isospin-dependent 
cross sections and Pauli blocking factors, the symmetry potential 
$V^{n(p)}_{{\rm asy}}(\rho,\beta)$ and Coulomb potential $V_c^p$ for protons. 
Besides $V^{n(p)}_{{\rm asy}}(\rho,\beta)$ and $V_c^p$ the nucleon mean-field 
$V^{n(p)}(\rho,\beta)$ also includes a symmetric term for which we use 
a Skyrme parameterization, i.e., 
\begin{equation}\label{pot}
      V^{n(p)}(\rho,\beta) = a (\rho/\rho_0) + b (\rho/\rho_0)^{\sigma}\ 
	+V^p_{c}+V_{{\rm asy}}^{n(p)}(\rho,\beta).
\end{equation}
In the above, the parameters $a,~b$ and $\sigma$ are determined by the 
saturation properties and the compressibility $K$ of symmetric nuclear 
matter\cite{bertsch}. The symmetric term should also contain
a momentum-dependent part. However, it is not essential for present study as we
will show that the ratio of preequilibrium neutrons to protons from 
collisions of neutron-rich nuclei is essentially independent of the 
symmetric part of the nuclear EOS. The isospin-dependent BUU model was 
used to explain successfully several isospin-dependent phenomena 
in heavy-ion collisions at intermediate energies\cite{lis95}. More recently,
the isospin-dependence of collective flow and balance energy 
predicted in ref.\ \cite{li96} using this model was confirmed 
experimentally at NSCL/MSU by Pak {\it et al.}\cite{pak1,pak2}. 

We have studied collisions of 
$^{112}Sn+^{112}Sn$, $^{124}Sn+^{124}Sn$ and $^{132}Sn+^{132}Sn$ 
reactions at a beam energy of 40 MeV/nucleon. The first two reactions 
have been recently studied experimentally at NSCL/MSU by the 
MSU-Rochester-Washington-Wisconsin collaboration\cite{kunde96,dem96}. 
Preequilibrium particles were measured
in these experiments and are now being analyzed\cite{udo}. The last system 
is included only for the purpose of discussions and comparisons. 
To identify free nucleons from those in clusters, we use in
our calculations a phase-space coalescence method at 200 fm/c after
the initial contact of the two nuclei, when the quadrupole moment of 
the nucleon momentum distribution in the heavy residue is almost zero, 
indicating the approach of thermal equilibrium.
A nucleon is considered as free if it is not 
correlated with other nucleons within a spacial distance of 
$\triangle r= 3$ fm and a momentum distance of $\triangle p = 300$ MeV/c.
We have checked that the results are not sensitive to these 
parameters if they are varied by less than 30\% around the above values.

\begin{figure}[htp]
\setlength{\epsfxsize=14truecm}
\centerline{\epsffile{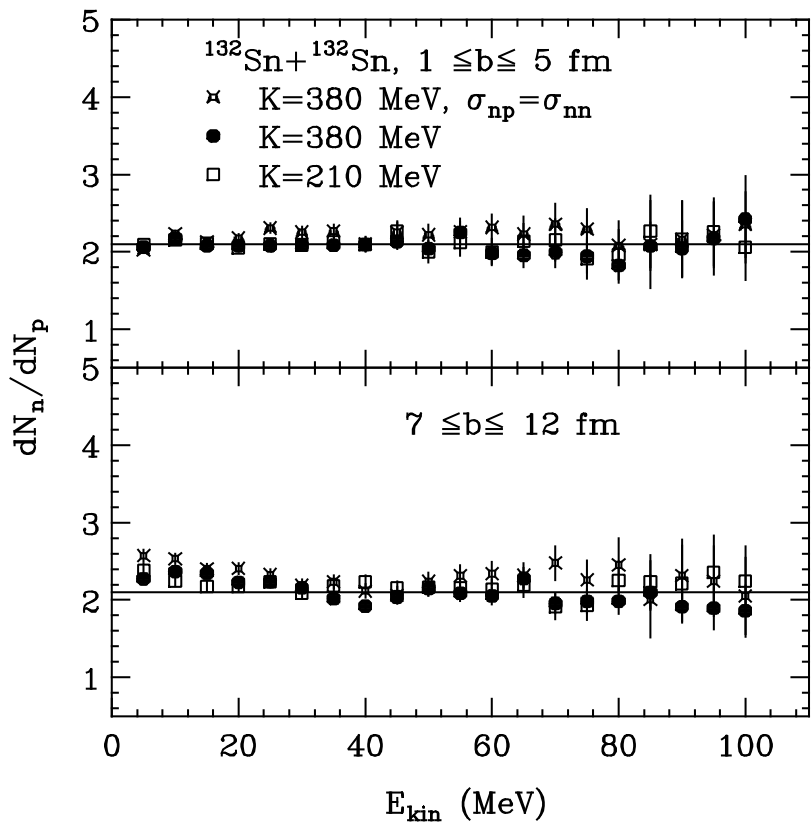}}
\caption{The ratio of preequilibrium neutrons to protons as a function
of nucleon kinetic energy for central (upper window) and peripheral 
(lower window) collisions calculated without the Coulomb and 
symmetry potentials.}
\label{ieosfig2} 
\end{figure}

We first study effects of the compressibility $K$ of symmetric 
nuclear matter and the in-medium nucleon-nucleon cross section on 
the ratio $R_{{\rm n/p}}(E_{{\rm kin}})$ by dropping both the Coulomb and 
symmetry potentials in the BUU model.
In Fig.\ \ref{ieosfig2} this ratio is shown as a function of 
nucleon kinetic energy for central (upper window) and 
peripheral (lower window) collisions of 
$^{132}Sn+^{132}Sn$ at a beam energy of 40 Mev/nucleon. 
When varying the compressibility $K$ from 210 MeV (open squares) to 
380 MeV (filled circles), we find that although the yields of both 
neutrons and protons increase, their ratio remains almost the same 
for all impact parameters. This is simply because the effects of symmetric 
EOS on both neutrons and protons are identical. 

The experimental cross section for neutron-proton 
collisions is about three times that for neutron-neutron (proton-proton) 
collisions in the energy range studied here. Setting the two cross 
sections equal (fancy squares), we find that the yields and 
their ratios change by less than 10\% even in peripheral collisions
of $^{132}Sn+^{132}Sn$. This result is also easy to understand since 
both colliding nucleons have the same probability to gain enough energy to 
become unbound\cite{lir}. Thus, the in-medium, isospin-dependent 
nucleon-nucleon cross sections do not affect much the ratio 
$R_{{\rm n/p}}(E_{{\rm kin}})$.
It is important to point out that in the absence of
Coulomb and symmetry potentials the ratios are almost independent of the
nucleon kinetic energy and have a constant value  
of about $2.1\pm 0.3$ in both central and peripheral collisions of
$^{132}Sn+^{132}Sn$. 

\begin{figure}[htp]
\setlength{\epsfxsize=14truecm}
\centerline{\epsffile{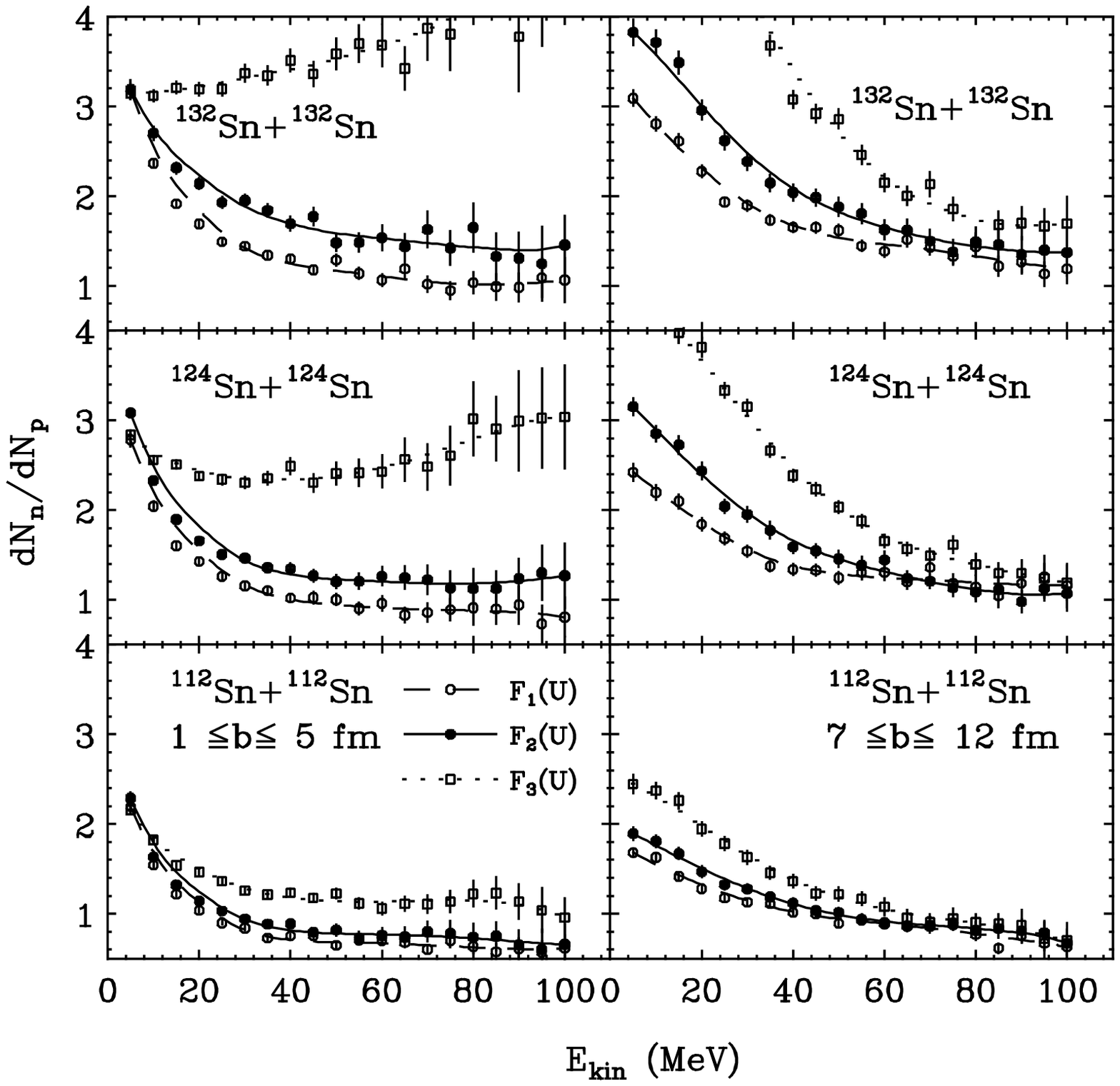}}
\caption{Same as in Fig.\ \protect\ref{ieosfig2} but calculated with 
both the Coulomb and symmetry potentials.}
\label{ieosfig3} 
\end{figure}

Including the Coulomb and the asymmetric term of the 
EOS in eq.\ (\ref{pot}), one can then study the effects of the symmetry
energy $S(\rho)$ since the Coulomb effect is well-known. 
We expect that the symmetry potential will have the following effects on 
preequilibrium nucleons. First, the symmetry potential 
$V_{{\rm asy}}^{n(p)}$ tends to make more neutrons than protons unbound.
One therefore expects that a stronger symmetry potential leads to 
a larger ratio of free neutrons to protons. 
Secondly, if both neutrons and protons are already free, the symmetry 
potential makes neutrons more energetic than protons. These 
effects are shown in Fig.\ \ref{ieosfig3} where we display the 
ratios calculated using the three forms of $F(u)$ for 
central (left windows) and peripheral (right windows) 
collisions of $^{112}Sn+^{112}Sn$, 
$^{124}Sn+^{124}Sn$ and $^{132}Sn+^{132}Sn$, respectively. 
The ratios change continuously from central to peripheral collisions, more 
detailed study on the impact parameter dependence of the ratio 
$R_{{\rm n/p}}(E_{{\rm kin}})$ will be published elsewhere\cite{rep}.
The increase of the ratios at lower kinetic energies in all cases 
is due to Coulomb repulsion which shifts protons from lower to 
higher kinetic energies. On the other hand, the different ratios 
calculated using different $F(u)$'s reflect clearly the 
effect mentioned above, i.e., with a stronger symmetry 
potential the ratio of preequilibrium neutrons to protons 
becomes larger for more neutron rich systems.

It is interesting to note that effects due to 
different symmetry potentials are seen in different kinetic energy 
regions for central and peripheral collisions. In central collisions, 
effects of the symmetry potential are most prominent at higher kinetic 
energies. This is because most of the finally observed free neutrons 
and protons are already unbound in the early stage of the reaction as 
a result of violent nucleon-nucleon collisions. The symmetry potential thus 
mainly affects the nucleon energy spectra by shifting 
more neutrons to higher kinetic energies with respect to protons. 
In peripheral collisions, however, there are fewer 
nucleon-nucleon collisions; whether a nucleon can become unbound
depends strongly on its potential energy. With a stronger symmetry potential 
more neutrons (protons) become unbound (bound) 
as a result of a stronger symmetry potential, but
they generally have smaller kinetic energies. Therefore, in peripheral 
collisions effects of the symmetry potential show up chiefly at lower 
kinetic energies. For the more neutron-rich systems the effects 
of the symmetry potential are so strong that in central (peripheral) 
collisions different forms of $F(u)$ can be clearly distinguished from 
the ratio of preequilibrium neutrons to protons at higher (lower) kinetic 
energies. However, because of energy thresholds of detectors, it is 
difficult to measure low energy nucleons, especially neutrons. Furthermore, 
the low energy spectrum also has appreciable contribution from equilibrium 
emissions. Therefore, the measurement of the ratio 
$R_{{\rm n/p}}(E_{{\rm kin}})$ in neutron-rich, central heavy-ion 
collisions for nucleons with energies higher than about 20 MeV is 
more suitable for extracting the EOS of asymmetric nuclear matter.
   
In conclusion, collisions of neutron-rich nuclei at intermediate energies 
reveal novel information about the EOS of asymmetric 
nuclear matter that is of interest to both nuclear- and 
astro-physics, such as the properties of radioactive nuclei, 
supernovae, and neutron stars. The ratio of the number of 
preequilibrium neutrons to that of protons is found to be sensitive 
to the asymmetric part, but not to the symmetric part of the nuclear EOS. 
It is also almost independent of the in-medium nucleon-nucleon cross sections. 
The experimental measurement of this ratio therefore provides a 
valuable means for determining the EOS of asymmetric nuclear matter.

We would like to thank J.B. Natowitz, W.Q. Shen and H.M. Xu 
for helpful discussions. This work was supported in part by the 
NSF Grant No. PHY-9509266. One of us (ZZR) was supported in part 
by grants from the Foundation of National Educational Commission 
of P.R. China.

\newpage
\newpage

\newpage

\newpage


\begin{thebibliography}{99}
\bibitem{tanihata95}I. Tanihata, Prog. of Part. and Nucl. Phys., 
{\bf 35} (1995) 505.

\bibitem{hansen95}P.G. Hansen, A.S. Jensen and B. Jonson, Ann. Rev. Nucl.
Part. Sci. {\bf 45}, 591 (1995).

\bibitem{sumi93}K. Sumiyoshi, D. Hirata, H. Toki and H. Sagawa,
	Nucl. Phys. {\bf A552}, 437 (1993).

\bibitem{toki95}K. Sumiyoshi, H. Suzuki and H. Toki,
	Astronomy and Astrophysics, {\bf 303}, 475 (1995).

\bibitem{na}T. Suzuki et al., Phys. Rev. Lett. {\bf 75}, 3241 (1995).

\bibitem{tanihata96}I. Tanihata, Preprint RIKEN-AF-NP-229, July, 1996.

\bibitem{lat91}J.M. Lattimer, C.J. Pethick, M. Prakash and P. Haensel,
		Phys. Rev. Lett. {\bf 66}, 2701 (1991).

\bibitem{sum94}K. Summiyoshi and H. Toki, 
	Astro. Phys. Journal, {\bf 422}, 700 (1994).

\bibitem{muller}H. M\"uller and B.D. Serot, Phys. Rev. C {\bf 52}, 2072 (1995).

\bibitem{kunde96}G.J. Kunde et al., 
	Phys. Rev. Lett. {\bf 77}, 2897 (1996).

\bibitem{dem96}J.F. Dempsey et al, 
	Phys. Rev. C{\bf 54}, 1710 (1996).

\bibitem{siemens70}P.J. Siemens, Nucl. Phys. {\bf A141}, 225 (1970).

\bibitem{baym71}G. Baym, H.A. Bethe, and C.J. Pethick, 
		Nucl. Phys. {\bf A175}, 225 (1971).

\bibitem{laga81}I.E. Lagaris and V.R. Pandharipande, 
	Nucl. Phys. {\bf A369}, 470 (1981)

\bibitem{wiringa88}R.B. Wiringa, V. Fiks and A. Fabrocini, 
		Phys. Rev. C{\bf 38}, 1010 (1988).

\bibitem{prak88}M. Prakash, T.L. Ainsworth and J.M. Lattimer, 
	Phys. Rev. Lett. {\bf 61}, 2518 (1988).

\bibitem{thor94}V. Thorsson, M. Prakash and J.M. Lattimer, 
		Nucl. Phys. {\bf A572}, 693 (1994). 

\bibitem{mass}P.E. Haustein, Atomic data and nuclear data tables, 
	{\bf 39}, 185-395 (1988).

\bibitem{farine}M. Farine, J.M. Pearson and B. Rouben, 
	Nucl. Phys. {\bf A304}, 317 (1978).

\bibitem{pear}J.M. Pearson et. al., Nucl. Phys. {\bf A528}, 1 (1991).

\bibitem{serot}B.D. Serot and J.D. Walecka, Adv. Nucl. Phys. {\bf 16}, 1 (1986).

\bibitem{glen82}N.K. Glendenning, Phys. Lett. {\bf B114}, 392 (1982).

\bibitem{rein88}P.-G. Reinhard, Z. Phys. {\bf A329}, 257 (1988).

\bibitem{rufa88}M. Rufa, P.-G. Reinhard, J. Maruhn, W. Greiner 
	and M.R. Strayer, Phys. Rev. C{\bf 38}, 390 (1988).

\bibitem{shar93}M.M. Sharma, M.A. Nagarajan and P. Ring, 
	Phys. Lett. {\bf B312}, 377 (1993).

\bibitem{toki94}Y. Sugahara and H. Toki, Nucl. Phys. {\bf A579}, 557 (1994).

\bibitem{scha96}J. Schaffner and I.N. Mishustin, 
	Phys. Rev. C{\bf 53}, 1416 (1996).
 
\bibitem{chin77}S.A. Chin, Ann. Phys. (N.Y.), {\bf 108}, 301 (1977).

\bibitem{horo87}C.J. Horowitz and B.D. Serot, 
Nucl. Phys. {\bf A464}, 613 (1987).

\bibitem{greiner}H. St\"ocker and W. Greiner, Phys. Rep. {\bf 137}, 277 (1986).

\bibitem{bertsch}G.F. Bertsch and S. Das Gupta, 
	Phys. Rep. {\bf 137}, 277 (1988).

\bibitem{libauer1}B.A. Li and W. Bauer, 
	Phys. Lett. {\bf B254}, 335 (1991); Phys. Rev. C{\bf 44}, 450 (1991). 

\bibitem{libauer2}B.A. Li, W. Bauer and G.F. Bertsch, 
	Phys. Rev. C{\bf 44}, 2095 (1991). 

\bibitem{ren}Z.Z. Ren, W. Mittig, B.Q. Chen, and Z.Y. Ma, 
	Phys. Rev. C{\bf 52}, R20 (1995).

\bibitem{lis95}B.A. Li and S.J. Yennello, Phys. Rev. C{\bf 52}, R1746 (1995).

\bibitem{li96}B.A. Li, Z.Z. Ren, C.M. Ko and S.J. Yennello, 
	Phys. Rev. Lett. {\bf 76}, 4492 (1996).

\bibitem{pak1}R. Pak et al., ``Isospin dependence of collective transverse 
flow in nuclear collisions'', Phys. Rev. Lett. (1997), in press.

\bibitem{pak2}R. Pak et al., ``Isospin dependence of the balance energy'',\\ 
Phys. Rev. Lett. (1997), in press.

\bibitem{udo}W.U. Schr\"oder, private communication.

\bibitem{lir}B.A. Li and J. Randrup, manuscript to be published. 

\bibitem{rep}B.A. Li et al., manuscript to be published.

\end{thebibliography}
\end{document}